\documentclass[number]{elsarticle}
\usepackage{latexsym}
\usepackage{natbib}
\usepackage[latin2]{inputenc}
\usepackage[hidelinks]{hyperref}
\usepackage{longtable, lineno}
\usepackage[final]{pdfpages}
\makeatletter
\def\ps@pprintTitle{%
	\let\@oddhead\@empty
	\let\@evenhead\@empty
	\def\@oddfoot{\centerline{\thepage}}%
	\let\@evenfoot\@oddfoot}
\makeatother

\begin{document}

\title{Comparing Advanced Graph-Theoretical Parameters of the Connectomes of the Lobes of the Human Brain}

%\linenumbers

\author[p]{Balázs Szalkai}
\ead{szalkai@pitgroup.org}
\author[p]{Bálint Varga}
\ead{balorkany@pitgroup.org}
\author[p,u]{Vince Grolmusz\corref{cor1}}
\ead{grolmusz@pitgroup.org}
\cortext[cor1]{Corresponding author}
\address[p]{PIT Bioinformatics Group, Eötvös University, H-1117 Budapest, Hungary}
\address[u]{Uratim Ltd., H-1118 Budapest, Hungary}

\date{}

%\linenumbers

\begin{abstract}
Deep, classical graph-theoretical parameters, like the size of the minimum vertex cover, the chromatic number, or the eigengap of the adjacency matrix of the graph were studied widely by mathematicians in the last century. Most researchers today study much simpler parameters of braingraphs or connectomes which were defined in the last twenty years for enormous networks -- like the graph of the World Wide Web -- with hundreds of millions of nodes. Since the connectomes, describing the connections of the human brain, typically contain several hundred vertices today, one can compute and analyze the much deeper, harder-to-compute classical graph parameters for these, relatively small graphs of the brain. This deeper approach has proven to be very successful in the comparison of the connectomes of the sexes in our earlier works: we have shown that graph parameters, deeply characterizing the graph connectivity are significantly better in women's connectomes than in men's. In the present contribution we compare numerous graph parameters in the three largest lobes --- frontal, parietal, temporal --- and in both hemispheres of the brain. We apply the diffusion weighted imaging data of 423 subjects of the NIH-funded Human Connectome Project, and present some findings, never described before, including that the right parietal lobe contains significantly more edges, has higher average degree, density, larger minimum vertex cover and Hoffman bound than the left parietal lobe. Similar advantages in the deep graph connectivity properties are hold for the left frontal vs. the right frontal and for the right temporal vs. the left temporal lobes.
\end{abstract}

\maketitle

\section*{Introduction} 

Structural connectomes, computed from diffusion weighted magnetic resonance imaging (MRI) data, are capable of describing the macroscopic connections between the anatomically identified small areas of the gray matter of the human brain. These connectomes can be viewed as mathematical objects called graphs: their nodes or vertices correspond to small gray matter regions of area 1.0-1.5 cm$^2$, frequently called ``Regions of Interests''(ROIs), and two such nodes are connected by an edge if a data-processing workflow finds axonal fibers in the white matter, connecting the ROIs of the nodes \cite{Daducci2012,Gerhard2011,Tournier2012,Fischl2012}. The number of nodes in these graphs are several hundred or around one thousand, and, typically, several thousand edges are identified between those nodes \cite{Hagmann2008}. 

One of the most frequently applied data sources for constructing connectomes is the public releases of the NIH-funded large Human Connectome Project (HCP) \cite{McNab2013}. Based on HCP data, we have computed and published hundreds of human connectomes, or braingraphs in GraphML format on the website \url{http://braingaph.org} \cite{Kerepesi2016b}. These macroscopic, anatomical connections can be examined as the macroscopic structural basis of all brain functions (e.g., \cite{Szalkai2016c}). 

An appealing research direction is the comparison of the braingraphs of individual subjects or groups of subjects with different biological/clinical parameters. It was shown on a publicly un-available dataset that the braingraphs of the sexes significantly differ \cite{Ingalhalikar2014b} in the inter-hemispheric/intra-hemispheric edge-number ratio. We have proven that women's connectomes not only differ, but have {\it much better} deep connectivity properties than that of men: female braingraphs have more edges, have larger bipartition width, larger minimum vertex cover, and are better expanders than that of men \cite{Szalkai2015,Szalkai2016a}. We have also shown that the women's advantage remains in effect if we compare large-brain females and small-brain males \cite{Szalkai2015c}; therefore the better connectivity properties of females are due to sex, and not some artifact, related to cerebral size or weight differences. The braingraph dataset applied in these works of us was computed from the high-quality HARDI (high angular resolution diffusion imaging) data of the Human Connectome Project \cite{McNab2013}, and it is publicly available at \url{http://braingaph.org} \cite{Kerepesi2016b}.

The comparison of the connectomes of the individual subjects -- instead of their groups -- is also a challenging question. The comparison of the individual braingraphs is possible: the set of the vertices of the braingraphs of all subjects is the same, since the brains of different shapes, size and weight can be mapped to the very same brain map, using the tools of FreeSurfer suite of programs \cite{Desikan2006,Fischl2012}. Consequently, the braingraphs of all subjects have the same, anatomically labeled  1015 nodes in the datasets examined in \cite{Kerepesi2015a, Szalkai2016, Szalkai2015a}, they differ only in the edges, running between some of the pairs of these nodes (c.f., Figure 1 of \cite{Szalkai2016}). 

For the description of the frequently appearing graph edges or connections, we have built the Budapest Reference Connectome Server \cite{Szalkai2015a, Szalkai2016}, available at the address \url{http://connectome.pitgroup.org}. The Server uses 477 connectomes, computed from Human Connectome Project's data \cite{McNab2013}, and the user may choose an integer $k$ between 1 and 477, and the server returns those edges which are present in at least $k$ connectomes of the subjects. Therefore, the frequently appearing edges can be mapped, while the infrequent connections can be filtered out. The graph of these $k$-frequent edges can also be visualized on the webpage \url{http://connectome.pitgroup.org}. 

The mapping of the individual differences of the cerebral connections could identify the more and less variable regions of the human brain. For the description of the variability we have used a natural mathematical tool, the cumulative distribution function in \cite{Kerepesi2015c}. We have shown, by analyzing the variabilities of the individual connectomes that some areas are more conservative (i.e., the variability of the graph edges is smaller) and some are more diverse (i.e., the variability of the graph edges are larger). For example, the limbic lobe is more conservative, while the edges in the temporal and occipital lobes are more diverse. Interestingly, a ``hybrid'' conservative and diverse distribution was found in the paracentral lobule and the fusiform gyrus of the human brain. Smaller cortical areas have shown also differences in connectome variability: the precentral gyri were found to be more conservative, and the postcentral and the superior-temporal gyri to be more diverse \cite{Kerepesi2015c}. 

In the present contribution we compare several advanced graph-theoretical parameters within the structural connectomes of the  largest three lobes of the human brain, the frontal, the temporal and the parietal lobes. We compute the parameters in both the left- and the right hemispheres in each of these lobes, and analyze the differences discovered. Only the largest lobes are analyzed graph-theoretically, since in the smaller lobes there are too few nodes to form graphs with rich enough structures.

\section*{Discussion and Results}

Here we define and explain the graph-theoretical parameters computed for the lobes in the current study. Some of the parameters are related to simple edge-counting (AvgDegree, Density, and Sum), some can be computed by more complex, but still polynomial-time \cite{Lovasz2007a} algorithms (AdjLMaxDivD, HoffmanBound, LogSpanningForestN, MinSpanningForest, PGEigengap) and some are known to be NP-hard \cite{Lovasz2007a}, and is computable by integer programming tools for small enough graphs (MinVertexCoverDivN, MinVertexCover). 

\subsection*{The definition of the braingraphs analyzed:} We have applied five different resolutions, namely 83, 129, 234, 463 and 1015 in the gray-matter parcellation for each subjects, applying the FreeSurfer tool \cite{Fischl2012}. Next, we have prepared five graphs for each subject, with 83, 129, 234, 463 and 1015 vertices, respectively, where the vertices correspond to the anatomically identified gray matter areas, and two vertices are connected by an edge if at least one neural fiber tract is identified between the corresponding two areas. 

It is important to make note the strength or the weight of the edge found between the pairs of vertices: some edges are defined by just one fiber tract, and others by dozens. For this goal we use five different weight functions for each edge: FAMean, FiberLengthMean, FiberN, FiberNDivLength and Unweighted, defined below. Therefore, for each subject, $5\cdot 5 = 25$ weighted graphs are computed and analyzed. 

We note that in most or, perhaps all connectomics studies, we are interested in the connections between the anatomically identified areas of the gray matter, while we do not want to follow or map the trajectories of the fiber tracts in the white matter. However, some physical properties of these trajectories (e.g., length, number of fibers) need to be recorded in the connectome: these properties are characterized by the different edge weights.

\subsection*{Weight functions on the edges}

The following weight functions are computed for the edges of the graphs:

FAMean: The fractional anisotropy  \cite{Basser2011}, averaged for the voxels on the tracts, defining the edge. For each voxel, the value of the fractional anisotropy is between 0 and 1: 0 if the diffusion-ellipsoid is a perfect sphere, and it is getting closer to 1 if the ellipsoid has one big and two small axes. If the average fractional anisotropy of a fiber tract is close to 1 then it means that most of the voxels of the fiber has an elongated diffusion ellipsoid. The reliability of the tractography in diffusion-weighted MR images mostly depends on the anisotropy of the voxels; consequently, the large FAMean value for an edge shows the reliability of the detected neural fiber tracts that define the edge.

FiberLengthMean: The average lengths of the fiber tracts, defining the graph edges, measured in mm.  

FiberN: The number of the fiber tracts, defining the edge in question. 

FiberNDivLength: The weight function FiberLengthMean does not reflect the number of the fibers; the weight function FiberN does not say anything about the lengths of the fibers, defining the edge. The FiberNDivLength weight function is the quotient of quantities FiberN and FiberLengthMean. It is an ``electric conductivity''-like graph parameter: it decreases for longer fibers and increases if multiple fibers define an edge. It can also be viewed as a reliability measure of the edge: with more detected fibers the reliability of the edge is increases, while with longer average length fibers the reliability decreases  \cite{Hagmann2008}. 

Unweighted: Evey detected edge has the same weight equal to 1.

\subsection*{Graph-theoretical parameters:} 

Most of the following graph-theoretical parameters were computed and analyzed for male and female connectomes in \cite{Szalkai2015,Szalkai2015c,Szalkai2016a}, and it was shown that women's connectomes have significantly better values for such parameters than those of men; here the ``better'' adjective refers to graph-theoretical and computer engineering measures of graph connectivity (e.g., \cite{Tarjan1983a}), as were detailed in \cite{Szalkai2015,Szalkai2015c,Szalkai2016a}, since the neuroscientific advantages of the ``better'' underlying structural network are not established yet.

Here we compute and compare these parameters in the frontal, parietal and temporal lobes of the human brain, both in the left and the right hemispheres. In what follows, we are dealing with the {\em induced subgraphs} of the human connectomes: the vertices correspond to the regions of interests (ROIs) in six cerebral areas: the left- and right frontal, temporal and parietal lobes, and only those edges are considered that connect two vertices from the very same area from these six one. In other words, edges that connect a vertex, for example, in the right frontal lobe to another vertex in the right parietal lobe are not considered; while all edges are considered that connect two vertices from the same area, say the right parietal lobe or the left parietal lobe, but edges that connect the left parietal lobe with the right parietal lobe are not considered. Consequently, we will made comparisons of the parameters of the graphs of these six cortical areas.

The vertex numbers of the six cortical areas in different resolutions are listed in Table 1. 

\begin{table}
\begin{tabular}{ | l | c | c | c | c | c | }
	\hline
	 & Scale 83 & Scale 129 & Scale 234 & Scale 463 & Scale 1015 \\ \hline
	Left\_Frontal & 10 & 19 & 37 & 74 & 168 \\ \hline
	Right\_Frontal & 10 & 18 & 36 & 74 & 167 \\ \hline
	Left\_Parietal & 6 & 13 & 30 & 62 & 132 \\ \hline
	Right\_Parietal & 6 & 13 & 29 & 60 & 136 \\ \hline
	Left\_Temporal & 7 & 10 & 18 & 33 & 74 \\ \hline
	Right\_Temporal & 7 & 10 & 17 & 33 & 74 \\ \hline
\end{tabular}
\caption{The vertex numbers in the six cerebral areas in different resolutions. We note that in all resolutions, the frontal lobes contain more nodes than the parietal and temporal ones, and except in the lowest resolution, the parietal lobes contain more vertices than the temporal ones.}
\end{table}

In the analysis below we list only the statistically significant ($p<0.05$) differences found in the graph parameters in the distinct lobes. The $p$ values mentioned are Holm-Bonferroni corrected ones \cite{Holm1979}, listed in Table S1 in the on-line supporting material. For the statistical details we refer to the "Statistical Analysis" subsection.

{\tt Sum:} In the Unweighted case, this parameter describes the number of the edges in each of the six areas of the cortex. For other weight functions (i.e., FAMean, FiberLengthMean, FiberN, FiberNDivLength), the {\tt Sum} parameter gives the sum of the weights of the edges in the area of question. We emphasize that the {\tt Sum} parameter and all the other parameters in this work are computed for the {\em induced graph edges} only: that is, both endpoints of the edges considered need to be in the same area from the six possible ones.

Figure S1 visualizes the Sum parameter in the Unweighted case (i.e., the edge number) in five resolutions and six areas. In all resolutions the larger lobes contain more edges than the smaller ones. The same area for the higher resolutions -- naturally -- contains more edges than in lower resolutions, simply because in higher resolutions there are more vertices in the graph. 

On Figure S1, in three resolutions out of five (83,129 and 1015 node resolutions), one can observe that the left frontal lobe contains significantly more edges than the right frontal lobe. The right parietal lobe contains significantly more edges than the left parietal lobe in three resolutions (129,234 and 1015 nodes) and the right temporal lobes, on average, contain significantly more connections than the left temporal lobes in resolution of 463 and 1015 nodes. 

The same significant advantage in the weighted number of connections in the case of the left frontal lobe and in the case of right parietal and temporal lobes hold true for Figures S2, S3, S4 and S5 with the four additional weight functions, and with most of the resolutions (see Table S1 for the corrected $p$ values for the different resolutions). For example, on Figure S3, the FiberN weighted edge numbers are significantly larger in the left frontal lobe in resolutions 234 and 1015, and the FiberN weighted edge numbers are significantly larger in the right 
parietal lobe than of the left in all resolutions. In the case of the temporal lobes, the right temporal lobe contains significantly more FiberN weighted edges than the left in all resolutions, except the 83-vertex one.

The observation concerning more connections in the left frontal lobe is surprising, since the width and the volume of the {\em right} frontal lobe is found to be larger on average than the left one \cite{Toga2003, Cherbuin2013}. The work \cite{Cherbuin2013}, based on structural and diffusion tensor MR imaging, has found also (in their Table II) that -- on the average -- the white mater of right frontal lobe is larger than of the left. Since axonal fibers (except the very short ones) are in the white matter, our finding seems to be interesting in the contrast of that result. 

Additionally, our finding of significantly more connections in the right temporal lobe needs also be compared to the results of lateral asymmetries described in \cite{Cherbuin2013}, Table II, where the size of the left temporal lobe is found to be larger in much more subjects (i.e., in 288 subjects), compared to cases where the right temporal lobe is larger (in 60 subjects). Consequently, we have found significantly more connections in the statistically smaller, right temporal lobe. Similarly to the data concerning the frontal lobe, the white matter of the left temporal lobe is found to be larger on average in \cite{Cherbuin2013}, while we have found more connections in the right temporal lobe. 

In the case of the parietal lobe, our results show significantly more connections within the right parietal lobe than in the left, while \cite{Cherbuin2013}, Table II also indicates larger volume in the right parietal gray matter and white matter than in the left ones.Therefore, the larger right-lobe volume coincides with more connections in the parietal lobe.

Another two parameters, {\tt AvgDegree} and {\tt Density} are closely related to the {\tt Sum} parameter. In the Unweighted case, the {\tt AvgDegree} parameter gives the average degree of the nodes in the induced subgraph: the degree of a node $v$ is the number of edges connected to $v$, the average degree is the sum of the degrees of the nodes, divided by the number of the nodes in the area considered. Again, we are counting the edges of the induced subgraphs, that is, both endpoints of the counted edge needs to be in the same area (out of the six considered) of the brain. Figures S6, S7, S8, S9 and S10 describe the {\tt AvgDegree} with different weight functions. 

In the unweighted case, for $n$ vertices, the {\tt Density} parameter describes the quotient $$ e\over{ {1\over 2}n(n-1)},$$ that is, the number of the edges (denoted by $e$) of the graph, divided by the maximum possible number (i.e., $n(n-1)/2$) of edges. For a complete graph, where each vertex-pair is connected by an edge ($e=n(n-1)/2$), this value is 1, for the empty graph (where no vertices are connected by edges) this value is 0. For the weighted case the Density is computed similarly, then $e$ should be substituted by the sum of the weights of the edges in the graph. Consequently, the density describes the relative total weight of connections in the area in question. 

Figure S11 shows the Density values in the Unweighted case, while Figures S12, S13, S14 and S15 the Density values with edge weights FiberNDivLength, FiberN, FiberLengthMean and FAMean, respectively. 

The Density values on Figure S11 are quite interesting. For rougher resolutions (blue and red bars) the density of the right and the left parietal lobes are the largest: in the 83 node resolution (blue bars) 80\% of the possible vertex-pairs are connected by edges, in the frontal and temporal areas these values are less than 60\%. However, in finer (i.e., higher) resolutions, the right and the left temporal lobes have the highest densities. This observation can be interpreted as follows: in low resolution a significantly larger fraction of the vertex-pairs are connected in the parietal area than in the other lobes, and in higher resolutions, relatively more vertex-pairs are connected in the temporal lobes than in the other lobes. Table S1 shows that the mentioned differences are all significant statistically. Similar relations can be observed for the Density with other weight functions on Figures S12, S13, S14 and S15.

The {\tt MinVertexCover} parameter is one of the most studied characteristics of graphs, examined for many decades (e.g., \cite{Gallai1959, GJ1979, Garey1974}). A vertex cover is a set of vertices in a graph that are incident to all edges in the graph (i.e., ``cover them''), see Figure 1. Clearly, the set of all vertices of any graph covers all the edges, so the question is not to find {\em any vertex cover}, but the minimum number of vertices that already cover all the edges; the parameter MinVertexCover is equal to this number. Clearly, the minimum vertex cover of the complete graph on $n$ vertices (where each pair of vertices are connected by an edge) is $n-1$ (one needs to include all vertices, except one; if two vertices were left out, the edge connecting them would not be covered), the minimum vertex cover of the empty graph on $n$ vertices (that contains no edges) has a MinVertexCover of 0, other graphs have this parameters between these two extreme values.
 
 \begin{figure} [h!]
 	\centering
 	\includegraphics[width=5in]{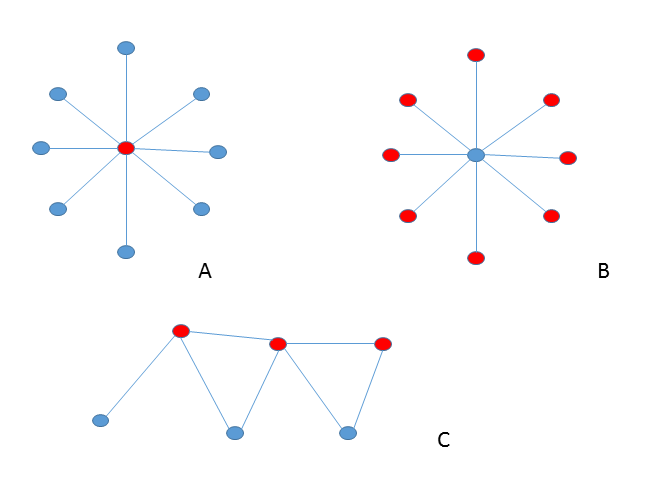}
 	\caption{Examples of the vertex covers of graphs. In these three graphs, the red nodes cover all the edges. In graph A, the central node is a minimum vertex cover. In graph B the red nodes are covering the edges, but they are not of minimal cardinality. In graph C the red vertices form also a minimum vertex cover. One can observe that the non-red vertices form independent vertex sets (they do not span any edges) in all three examples: this is not by chance, the complementer vertex-set of all vertex covers are independent sets, they cannot span edges, otherwise, the spanned edges were not covered by the vertex cover. Therefore, the complementer set of the minimum vertex cover needs to be the maximum independent set in any graph. Consequently, finding the minimum vertex cover is equivalent to the finding the maximum independent vertex set in a graph. }
 \end{figure}
 
 As one can easily verify (e.g., on Figure 1), the complement set of any vertex cover cannot span any edge (since this edge would not be covered), therefore, this set is an independent set (a vertex set is called to be an independent set, if it does not span any edges). Consequently, the complement of the minimum vertex cover set is the maximum independent set in a graph, so, if we denote the size of the minimum vertex cover set of a graph $G$ by $\tau(G)$, and the size of the largest independent set in $G$ by $\alpha(G)$, then, for an $n$-vertex graph $G$:
 $$\tau(G)+\alpha(G)=n.$$ 
 It is well known that computing $\tau(G)$ (and, equivalently, $\alpha(G)=n-\tau(G)$), is an NP-hard problem \cite{GJ1979, Garey1974}, therefore, it is not probable that it can be computed by a fast (i.e., polynomial time) algorithm. One can interpret, in a sense, these parameters as follows: dense, well-connected graphs have a larger MinVertexCover and, consequently, a smaller independence number ($\alpha(G)$), while "scarcely" connected graphs have a smaller  MinVertexCover and, consequently, a larger independence number. Apart from the independence number, the minimum vertex cover is also closely related to the matching number of the graphs \cite{Gallai1959}.
 
 Figures S16 through S20 depict the MinVertexCover values for different lobes. Generally, the higher resolutions, with larger vertex numbers (Table 1) imply larger MinVertexCover values. On Figures S17 and S18, where the weights are proportional to the fiber numbers, there are no such increases in finer resolutions, since the same fiber-set is re-partitioned in finer resolutions in the graph construction process (see the Methods section).
 On Figure S16, the Unweighted case is covered: in higher resolutions (129, 234 and 1015 vertices), the already observed differences are strengthened between the left and right frontal, parietal and temporal lobes. These results show that the left frontal lobe contains not just more connections than the right frontal lobe, but its minimum vertex cover is also significantly larger (in the 463-vertex resolution the difference is not significant). 
 
 We need to note that more edges do not necessarily imply larger MinVertexCover values as the example on Figure 2 clearly demonstrates. 
 
 \begin{figure} [h!]
 	\centering
 	\includegraphics[width=5in]{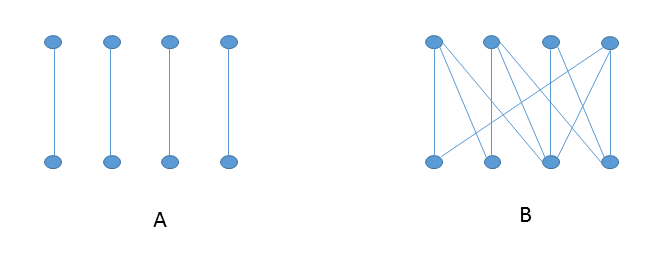}
 	\caption{Graph A and B have the same MinVertexCover value (i.e., 4), while graph A has four edges and graph B 11 edges. Therefore, more edges do not necessarily imply larger MinVertexCover value.}
 \end{figure}

{\tt MinVertexCoverDivN:} It can be seen clearly on Figure S16 (and also on Figures S19 and S20) that in all six areas of the brain the MinVertexCover values are increased along the increase of the vertex number (or the resolution). Therefore, it would be interesting to compute the differences in the {\em relative} size of the vertex covers, compared to the number of the vertices in the area. The parameter MinVertexCoverDivN describes exactly this value for each area under study: we normalized by dividing the MinVertexCover parameter by the vertex number of the lobe, in each resolutions. 

The average of the normalized vertex cover numbers are visualized on Figures S21-S25. Figure S21 is particularly informative in the comparison of the lobes: now the vertex-number differences are ``factored out'', because of the normalization, and we can compare the parameters of the different lobes, without the influence of the differing vertex numbers. 

One can observe that in the higher resolutions the MinVertexCoverDivN values are significantly higher in the frontal lobe than in the parietal and in the temporal lobe, and it is larger in the parietal lobe than in the temporal one. Therefore, in this sense --- and strictly in this sense --- we can conclude that the frontal lobe has more ``complex'' graph than the parietal and the temporal, and the parietal has a more ``complex'' graph than the temporal lobe. We note that -- in the higher resolutions -- the advantage of the left frontal lobe against the right frontal lobe, and the advantage of the right parietal and the temporal lobes against the left ones remained significant, even in this vertex-number normalized setting. Similar observation can be made on Figure S25, with the FAMean weight function, except that the higher value in the left frontal lobe only in scale 463 is significant.

{\tt PGEigengap:}  The transition matrix $P_G$ of a graph $G$ describes the Markovian process of the random walk on the graph, i.e., one walker visits the vertices of the graph in a sequence of steps as follows: when the walker is in a node then he makes a (uniform) random choice between the adjoining edges, and follows that edge to the other endpoint. When the graph is weighted by positive weights then the probability of choosing an edge is proportional to its weight. Matrix $P_G$ contains the probabilities of this walk. The eigengap of matrix $P_G$ is the difference between its largest and second largest eigenvalues, and this value is characteristic to the expanding properties of the graph \cite{Hoory2006}: a larger eigengap implies a better expander graph. 

Expander graphs with few edges are important objects of study in mathematics and in engineering, and have dozens of important applications. Let us consider $n$ vertices, and the $n$-vertex complete graph, denoted by $K_n$: in this graph every pair of vertices are connected by an edge, therefore, $K_n$ has $n(n-1)/2$ edges. $K_n$ has perfect connectivity properties: from each vertex each other vertex can be reached in a short path (in fact, through a single edge), if we make a short random walk then the probability of being at any vertex will be almost the same (in fact, already after the first step), or any two, disjoint sets, containing $k$ vertices each, are joined by $k$ vertex-disjoint paths (in fact, by $k$ edges). Consequently, the graph $K_n$ has very good connectivity properties. However, it has a great disadvantage: it has too many edges ($n(n-1)/2$ edges), and each vertex has a too large degree ($n-1$). For example, if the human brain's $n=80$ billion neurons were formed the vertices of a complete $n$-vertex graph, $K_n$, then each neuron needs to be connected to 80 billion others; that is, the small vicinity of each neuron needs to contain 80 billion connecting axons! This is simply impossible, there is no place for so many axons in the small vicinity of each neurons.

Consequently, mathematicians and engineers are interested in graphs with {\em very good} connectivity properties, similar to that of the complete $n$-vertex graph, but with {\em very few} edges and only a small degree at each nodes. These small-degree expander graphs are of the subjects of an intensive research for many decades now \cite{Hoory2006}. 

We have computed the eigengap of the $P_G$ transition matrix for all lobes of all subjects, and their average value is denoted by PGEigengap. The results are visualized on Figures S26 through S30.

On Figure S26 the PGEigengap with Unweighted edges are demonstrated. In finer resolutions, the PGEigengap is larger in the frontal and in the temporal lobes than in the parietal lobe, and it is the highest in the temporal lobe. In other weight functions (Figures S27-S30) the advantage of the temporal lobe remains true. In finer resolutions, the differences between the left and right parts of the same lobes are not significant in most weight functions.

{\tt AdjLMaxDivD:} The adjacency matrix $A$ of an $n$-vertex graph is an $n\times n$ matrix, where the rows and the columns are corresponded to the vertices of the graph. The entry in row $i$ and column $j$ is 1 if vertices $v_i$ and $v_j$ form an edge, and 0 otherwise. In the case of weighted graphs, the entry in row $i$ and column $j$ is the $w_{ij}$ weight of the $\{v_i,v_j\}$ edge, or 0 if no such edge exists. For un-weighted  non-directed graphs the eigenvalue with the largest absolute value of the adjacency matrix $A$ is a non-negative real number $\lambda$ (because of the Perron-Frobenius theorem \cite{Lovasz2007,Hoffman1972}), and its value is larger or equal to the average degree of the vertices and less or equal to the maximum degree of the vertices \cite{Chung1997}. For non-empty graphs, {AdjLMaxDivD} denotes the value $\lambda/d$, where $d$ denotes the average degree of the vertices. For unweighted graphs, its value is always at least 1. For weighted graphs, the degrees of the vertices need to be computed by summing the weights of the adjoining edges.

In regular graphs (where all the degrees are the same number, say $d$), the maximum and the average degrees coincide, therefore $\lambda=d$, and AdjLMaxDivD is 1. Consequently, in a sense, AdjLMaxDivD describes the ``irregularity'' of the degree distribution of the graph: the higher the number the more ``irregular'' the degree distribution. 

Figures S31-S35 visualizes the AdjLMaxDivD values for different weight functions. On Figure S31, in higher resolutions, we can learn that the ``more complex'' left frontal lobe is more regular in this sense than the  ``less complex'' right frontal lobe (for 1015-vertex resolution) and the graph of the ``less complex'' left temporal lobe is more irregular than the ``more complex'' right temporal lobe (in 463 and 1015 vertex resolutions). In the case of the parietal lobe the differences exist, but they are not significant statistically. On Figure S35, with the FAMean weight function, the left parietal lobe is more ``irregular'' than the more ``complex'' right parietal lobe (in the 129, 234 and 463 node resolutions the differences are significant).

{\tt HoffmanBound:} This value gives a lower bound to the (hard-to-compute) chromatic number of graphs. The chromatic number of graphs is the smallest integer $k$ that the vertices of the graph can be colored by $k$ colors, in a way that adjoint vertices carry different colors. The complete $n$-vertex graph clearly needs different colors for all of its vertices, while, for example, any tree can be colored by two colors. In a certain sense, the graph chromatic number can also be applied as a measure of graph complexity: more complex graphs have higher chromatic numbers. We note that the chromatic number is rather independent from the edge number of the graph: for example, the $n$-vertex bipartite graph with two equal vertex classes has $n^2/4$ edges and its chromatic number is only 2, while an $n$-vertex graph that consists of a single $[\sqrt{n}]$-clique has less than $n$ edges and its chromatic number is high: $[\sqrt{n}]$. In this sense we apply the Hoffman bound also as a complexity measure; it is defined as
$$1 + \frac{\lambda_{max}}{|\lambda_{min}|},$$ where $\lambda_{max}$ and $\lambda_{min}$ denote the largest and smallest eigenvalues of the adjacency matrix of the graph \cite{Hoffman1972}. 

The Hoffman bounds are visualized on Figures S36-S40. On Figure S36, in the Unweighted graphs, in higher resolutions, the right parietal and temporal lobes have significantly larger HoffmanBounds than the left ones, and in scales 129 and 463, the left frontal lobe has larger HoffmanBound than the right one. 

The {\tt LogSpanningForestN} parameter describes the logarithm of the number of the spanning forests in the unweighted case. This parameter can also be applied to describe the ``connectedness'' of the graph, in the sense, detailed below. We say that a graph $G$ is connected, if from any vertex $u$ any other vertex $v$ can be reached on a path, consisting of a sequence of edges of the graph. A tree is a minimal connected graph: deleting any edge would make the graph non-connected. Any tree on $n$ vertices contains exactly $n-1$ edges. For a connected graph $G$ $T$ is a spanning tree of $G$ if $T$ contains all the vertices of $G$, and $T$ is a tree subgraph of $G$. 

Clearly, all connected graphs contain at least one spanning tree. A tree (as the minimal connected graph) has only one spanning tree: itself. One may quantify the ``connectedness'' of a connected graph $G$ by the number of spanning trees of $G$. Cayley's theorem states that the $n$-vertex complete graph has $n^{n-2}$ spanning trees. Any connected, $n$-vertex $G$ has at least 1 and at most $n^{n-2}$ spanning trees. Computing the number of the spanning trees in a connected $G$ can be done from the eigenvalues of $G$'s Laplacian matrix, by applying the famous matrix-tree theorem of Kirchoff, proved in 1847 \cite{Kirchoff1847}. 

If $G$ is not connected, then, instead of spanning trees, it has spanning forests. The parameter  LogSpanningForestN gives the logarithm of the number of the spanning forests in the unweighted case. When edge weights are present, the parameter LogSpanningForestN gives the sum of the logarithms of the added-up edge-weights of trees in the forest. This value can also be negative if all the weights are small.

Figures S41-S45 demonstrate the LogSpanningForestN parameters with different weight functions. Figure S41 (the Unweighted case) shows that in the 1015-vertex resolution, this parameter is larger in the left frontal lobe than in the right, and in the right parietal and in the right temporal lobes than in the left ones. Similar observations can be made on Figures S43 and S44. 

The {\tt MinSpanningForest} parameter gives the average weight of the minimum spanning forests for each lobe. For connected graphs, the minimum-weight spanning tree is an extremely important structure with lots of applications. In the unweighted case (more exactly, when each edge-weight is 1), if the $n$-vertex graph $G$ is connected, then the weight of its minimum-weight spanning tree is always exactly $n-1$. If graph $G$ has $\ell$ connected components, then the weight of its minimum-cost spanning forest is exactly $n-\ell$. Therefore, in the unweighted case, the MinSpanningForest parameter describes the component number of $G$. 

In weighted case, its value is the sum of the weights of the minimum weight spanning trees in each of its components. Figures S46-S50 describe this value with different weight functions, while Figures S51-S55 describe this value normalized by the vertex number minus one (i.e., the edge number of the tree) in each lobe (denoted by MinSpanningForestDivNMin1).

Figure S51 shows that in the left parietal lobe in the coarsest two, and in the right parietal lobe in the coarsest three resolutions the graphs are almost always connected (i.e., their component number is 1). In finer resolutions the number of the connected components are typically larger, that is, the ``connectivity'' is, in this sense, worse. The relative number of graph components is the smallest in the 234-vertex scale in the frontal and the temporal lobes. This observation is paralleled by the Density parameter values on Figure S11, where similar observations are done.

\section*{Methods}

\subsection*{Computing the connectomes:} We have applied the graphs from the \url{braingraph.org} repository \cite{Kerepesi2016b}, which were computed from the Human Connectome Project's public data release \cite{McNab2013}. The details of their construction is given in \cite{Kerepesi2016b}. 

\subsection*{The computation of the graph-theoretical parameters:} The graph-theoretical parameters, detailed in the ``Discussion and Results'' section, were computed for the spanned subgraphs for the left- and right frontal, parietal and temporal lobes. That means that only those edges were allowed, which have both of their endpoints in the very same area from these six ones. The subgraphs were prepared from the \url{braingraph.org} data; in this work, we have analyzed the graphs of 423 subjects from the Human Connectome Project \cite{McNab2013}. The parameters were computed by our own scripts and by the integer programming (IP) solver SCIP \url{http://scip.zib.de}, \cite{Achterberg2008, Achterberg2009}.

\subsection*{Statistical analysis:} For each graph parameter, resolution and weight function, we applied a statistical null-hypothesis \cite{Hoel1984} that the average values of the graph parameters do not differ for distinct lobes. Next, we applied Holm-Bonferroni \cite{Holm1979} corrected ANOVA (Analysis of variance) \cite{Wonnacott1972} for assigning p-values for each pairs of values with the same parameter, resolution and weight function.  The null hypothesis was refuted if the corresponding corrected p-value was less than 0.05. 

\section*{Conclusions}

We have analyzed numerous deep graph-theoretical properties in the left- and right frontal, parietal and temporal lobes of the connectomes of 423 subjects, computed from the data of the Human Connectome Project \cite{McNab2013}. The most relevant findings of our analysis are listed as follows:

\begin{itemize}
\item In numerous resolutions, the left frontal lobe contains significantly more edges, has higher average degree, density, larger minimum vertex cover, Hoffman bound and has a ``more regular'' degree distribution than the right frontal lobe.

\item In numerous resolutions, the right parietal lobe contains significantly more edges, has higher average degree, density, larger minimum vertex cover and Hoffman bound than the left parietal lobe.

\item In numerous resolutions, the right temporal lobe contains significantly more edges, has higher average degree, density, larger minimum vertex cover and Hoffman bound and has a ``more regular'' degree distribution than the left temporal lobe.

\item Comparing the normalized MinVertexCoverDivN parameter in finer resolutions in frontal, parietal and temporal lobes, one can conclude that these values are largest in the frontal and the smallest in the temporal lobes. In this sense, the frontal lobe is more complex than the other two lobes.

\item The eigengap differences show that in the coarsest resolution the parietal lobe is the best expander, and in the finer resolutions the temporal lobe is the best expander. In the case of the temporal lobe, this remark is in line with the density-differences: in the finest three resolutions the temporal lobe has the highest densities. In this sense, the temporal lobe is more ``well-connected'' than the other two ones.

\item Both the left and right parietal lobes are almost always connected in the coarsest three resolutions, for the frontal and temporal lobes this property does not hold.
	
\end{itemize}

\section*{Data availability:} The data source of this study is Human Connectome Project's website at \url{http://www.humanconnectome.org/documentation/S500} \cite{McNab2013}. The connectomes that we have computed from these data are freely available at the site \url{http://braingraph.org/download-pit-group-connectomes/} \cite{Kerepesi2016b}. The Supporting figures are enclosed with this work, after the Reference section. The statistical details (average parameter values, standard deviations, and the corrected p-values) are available as a very large Table S1 from \url{http://uratim.com/paralobe/Table_S1.pdf}.

\section*{Acknowledgments}
Data were provided in part by the Human Connectome Project, WU-Minn Consortium (Principal Investigators: David Van Essen and Kamil Ugurbil; 1U54MH091657) funded by the 16 NIH Institutes and Centers that support the NIH Blueprint for Neuroscience Research; and by the McDonnell Center for Systems Neuroscience at Washington University. 
The authors declare no conflicts of interest.

%\bibliography{v:/vince/CIKKEK/medl}
%\bibliographystyle{unsrtnat}

\setboolean{@twoside}{false}

\includepdf[pages=-, offset=0 0]{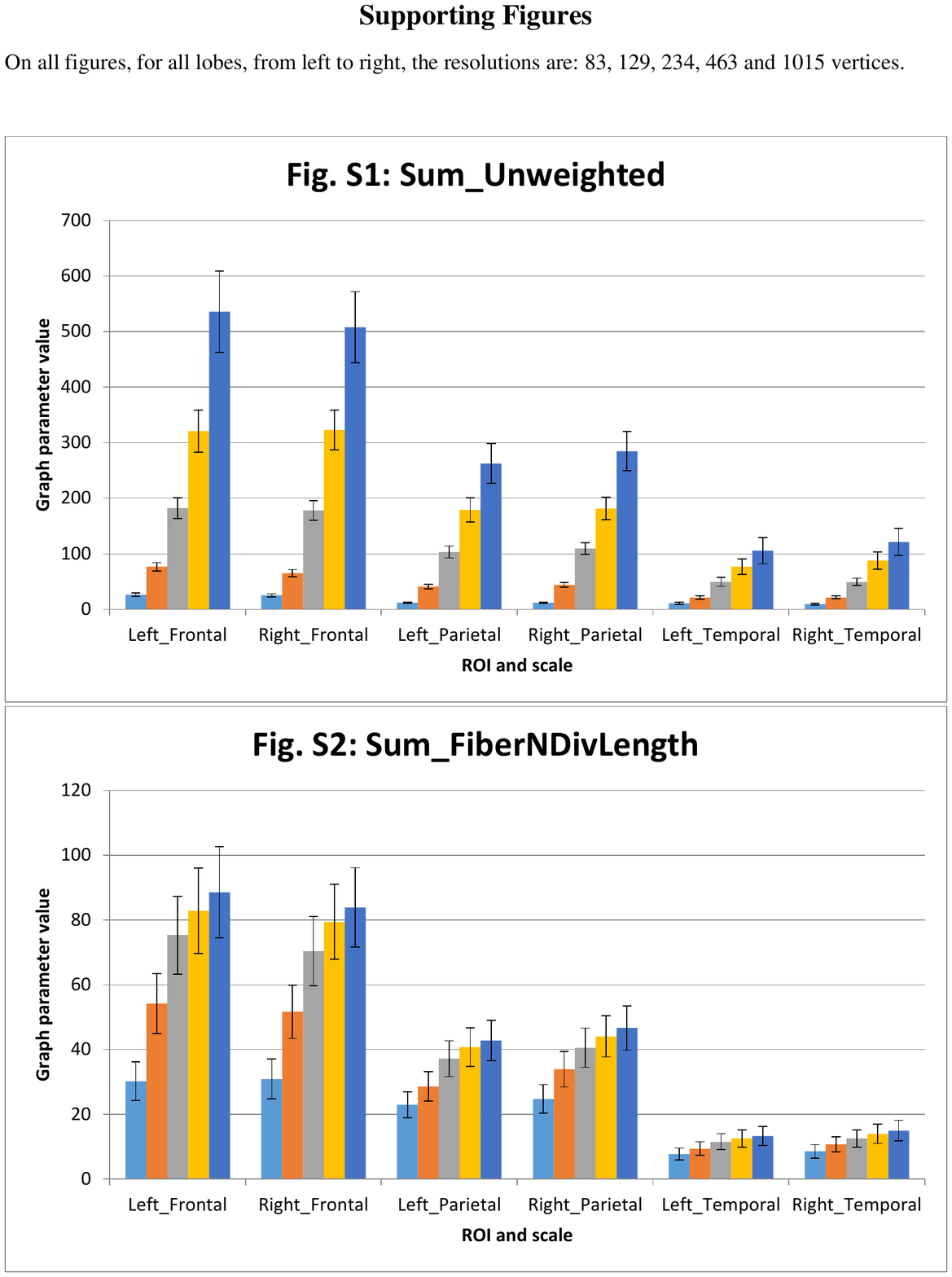}

\end{document}